# Interatomic potential theory on the phase transition of charge density wave in transition metal dichalcogenides

Changwon Park*

*School of Computational Sciences, Korea Institute for Advanced Study, Hoegiro 85, Seoul 02455, Korea*

## Abstract

Patterns and periods of charge density waves (CDW) in transition metal dichalcogenides exhibit complex phase diagrams that depend on pressure, temperature, metal intercalation, or chalcogen alloying. The phase diagrams have been understood in the context of Landau free energy model, but available methods to calculate many parameters in the model are still lacking. Here, we present that the total energy function based on the interatomic potential has similar structure with the model and the parameters correspond to effective force constants. Using the eigenmode representation of the interatomic potential extracted from first-principles calculations, we have explicitly calculated a temperature-dependent phase diagram of monolayer H-TaSe$_2$, where the followings are manifested: 1) commensurate lock-in, 2) stripe CDW phase, and 3) commensurate-commensurate phase transition. Our work shows the complex behaviors of charge density wave are originated from the relatively simple structure of the interatomic potential and elucidates the role of lattice anharmonicity on the CDW phase transition.

## I. INTRODUCTION

The charge density wave (CDW) is a spatial ordering of charge density that typically occurs in conjunction with a periodic lattice modulation. The new periodicity introduced in the crystal not only modifies its electrical properties, but it also competes with the existing lattice periodicity, resulting in a variety of interesting phenomena as plasticity [1] and superlubricity [2,3]. Charge density waves in transition metal dichalcogenides (TMDCs) are of particular interest because they are quite ubiquitous [4-8] and have well-defined phase diagrams that depend on pressure, temperature, metal intercalation or chalcogen alloying [9-13]. Theoretical descriptions of CDW phase transition are pioneered by McMillan [14,15], Nakanishi and Shiba [16] based on the parametrized Landau free energy models. Many aspects of CDW phase transitions such as commensurate lock-in and discommensuration [15] can be well reproduced in their models by the two competing energy terms favoring commensurate and incommensurate period, respectively.

Motivated by the success, there has been many approaches constructing the free energy function in an *ab initio* manner. Scenarios based on the Fermi surface nesting or singularities in density of states [17] emphasized the role of electron-electron interaction as a driving force for CDW, but it has been widely accepted that both electron-electron and electron-phonon [18] interactions are equally important to reconcile the absence of diverging electron susceptibility and CDW formation in some CDW materials [19, 20]. The importance of phonon fluctuation is also demonstrated in the quantitative evaluation of critical temperature of CDW [21].

The free energy function can be directly obtained from first-principles calculations in terms of interatomic potential. In fact, the approach is already widely used for structural phase transitions in ferroelectric materials [22,23] and has been successful to describe the commensurate phase of CDW because there is nothing to distinguish a CDW from a structural phase transition in real materials [24]. When we apply the interatomic potential approach to CDW phase transitions, we encounter two practical difficulties: 1) the long-range nature of the force constant being inherent in metals, and 2) the incommensurate CDW phase demanding huge supercell calculations. To overcome these difficulties, we developed a practical method calculating the interatomic potential energy in eigenmode representation.

The method is applied to calculate the temperature-dependent phase diagram of CDWs in monolayer hexagonal phase of transition metal dichalcogenides (H-TMDC), which are known to have either 3×3 commensurate CDW (CCDW) or slightly incommensurate CDW (ICDW). The interatomic potential is extracted from density functional theory calculations of $3 \times 3$ supercell and density functional perturbation theory (DFPT) [25] without any adjustable parameters. Finite temperature effect is incorporated in the interatomic potential in a mean-field manner. A few natural conditions for minimum structure to meet are derived from the eigenmode representation of the interatomic potential, and we can construct candidate CCDW and ICDW structures in variational forms. By optimizing the temperature-dependent potential energies of CDWs and comparing their energies, we can obtain the temperature-dependent phase diagram with the following features: 1) commensurate lock-in, 2) stripe CDW phase [26], and 3) commensurate-commensurate phase transition. We also show that two competing commensurate and incommensurate periods naturally emerge from anharmonic and harmonic part of the potential, and the complex behaviors of charge density wave are originated from the relatively simple structure of the interatomic potential.

## II. METHODS

### A. Eigenmode representation of charge density wave

The atomic displacements of a CDW can be represented with eigenmodes of the crystal. Here, eigenmodes are principal axes of second-order (harmonic) part of the potential, or equivalently, eigenvectors of dynamical matrix without mass normalizations. Because we are considering a static displacement, not a vibration, two eigenmodes with wavevectors $\pm \vec{k}$ should be complex conjugates of each other and a static phase of the eigenmode should be considered. By using an eigenmode representation, we can effectively reduce the atomic degrees of freedom because constituting eigenmodes of CDW is mostly from the lowest branches (From our density functional theory calculations, for commensurate CDW of H-TMDC's, over 90 % of displacements are projected on the lowest branches.).

The eigenmodes can be calculated from standard methods for phonon calculation when the displacements are small. But atomic displacements in CDW are rather large (~0.1 Å for H-TaSe$_2$) that eigenmodes obtained from phonon calculations can result in large errors in the potential energy. Instead, we used the property that for a pure harmonic potential, the lowest-branch eigenmodes are variational minima of the interatomic potential if their amplitudes and phases are fixed. In the presence of anharmonic potential, the eigenmodes slightly depends on the of amplitudes and phase constraints that appropriate average procedures over various constraints are required. For high-symmetric eigenmodes, we can set parametrized forms of them considering the lattice symmetry, and optimize their energy to get the parameters.

## B. Eigenmode representation of interatomic potential

The interatomic potential is defined as the energy required for displacing atoms from a reference configuration. Practically, the potential is expanded with a Taylor series of atomic displacements and the reference configuration is set to be a saddle point or local minimum of the potential. By a change of variables from atomic displacement to eigenmodes, the interatomic potential becomes a function of amplitudes and phases of eigenmodes. When $n$ eigenmodes of amplitudes $A_i$ and phase $\theta_i$ are given, the interatomic potential can be generally written in ascending order of amplitudes as

$$E = \sum_i \varepsilon_i A_i^2 + \sum_{i,j,k} \sum_{s_2,s_3} f_{ijk}(s_2,s_3) c_{ijk} A_i A_j A_k \cos(\theta_i + s_2\theta_j + s_3\theta_k + \gamma_{ijk}) + \cdots \quad (1)$$

where $\epsilon_i$ is the energy of an eigenmode, $s_2$ and $s_3$ are either +1 or -1, and $c_{ijk}$ and $\gamma_{ijk}$ are a coefficient and phase constant determined from the force constants and polarization vectors of eigenmodes. $f_{ijk} = 1$ when the following wavevector relation

$$\vec{k_i} + s_2\vec{k_j} + s_3\vec{k_k} = n\vec{G} \quad (2)$$

holds, or $f_{ijk} = 0$. The derivation of Eq. (1) can be found on Appendix A.

The first term on the right side of Eq. (1) is the energies of eigenmodes and the following terms can be understood as their interactions including self. For example, the higher-order potential energy of two eigenmodes whose wavevectors $\vec{k_1}$ and $\vec{k_2}$ satisfy $2\vec{k_1} + \vec{k_2} = \vec{G}$, can be calculated as follows. The trivial relations 1) $2\vec{k_1} - 2\vec{k_1} = 0$ and 2) $\vec{k_1} - \vec{k_1} + \vec{k_2} - \vec{k_2} = 0$ result in phase-independent fourth-order energy terms being proportional to $A_1^2 A_1^2 = A_1^4$ and $A_1^2 A_2^2$, while the nontrivial relation $2\vec{k_1} + \vec{k_2} = \vec{G}$ results in a third-order energy term of the form $\alpha A_1^2 A_2 \cos(2\theta_1 + \theta_2 + \beta)$ for material-dependent parameters $\alpha$ and $\beta$.

For eigenmodes with commensurate wavevectors, the coefficients and phase constants can be extracted using Eq. (1) as a fitting function to reference density functional theory calculations. For eigenmodes with slightly incommensurate wavevectors, it is impossible to perform reference calculations that we used DFPT to calculate $\varepsilon_i$ of those eigenmodes. Because DFPT is accurate only when the amplitude of eigenmode is small, calculated $\varepsilon_i$ is not the same with $\varepsilon_i$ obtained with the above fitting method. To match their $\varepsilon_i$'s at commensurate wavevectors, we introduced an overall scaling factor for DFPT results. For the higher-order potential coefficients and phase constants, we approximated them with those of the closest commensurate ones (zeroth-order approximation). When two eigenmodes have similar wavevectors, integer factor should be multiplied as derived in Appendix B.

Our density functional theory calculations employ the Perdew–Burke–Ernzerhof exchange-correlation functional [27] and the projector augmented wave method for ionic potentials [28] as implemented in the Vienna Ab Initio Simulation Package (VASP) [29]. The energy cutoff of planewave basis is 400 eV and 18×18 k-points are sampled for the unitcell (equivalent density for supercells) of 1H-TMDC's. For the phonon calculation of 1H-TaSe$_2$, density functional perturbations theory (DFPT) [25] implemented in Quantum Espresso [30] are adopted. The energy cutoff of planewave basis is 50 Ry and 24×24 k-points are sampled for the real space force constants.

### C. Temperature-dependent effective potential

The effect of temperature can be incorporated to the lattice potential with a simple prescription increasing the short-range repulsion. The rationale for the prescription is as follows. Thermal fluctuations of atoms result in changes in instantaneous distances between atoms, and we can define the "average" potential energy between them. If the interatomic potential is truncated at the fourth-order, the change in distances results in an energy cost from the (positive) fourth-order repulsive potential in the form of second-order polynomial. Though the third-order potential makes some energy gain from a dynamical correlation between atoms, it plays no role within a mean-field level. More rigorous arguments and an example for one-dimensional atomic chain is presented in Appendix C.

## III. RESULTS AND DISCUSSIONS

### A. Eigenmodes and interatomic potential of monolayer H-TMDC

In our approach, CDW is the optimized atomic coordinate for a given interatomic potential. In eigenmode representation, Eq. (1) should be optimized about wavevector, amplitudes, and phase of eigenmodes. Because Eq. (1) has infinitely many variables (amplitudes and phases of eigenmodes), we should first restrict the number of eigenmodes. Let's first consider 3×3 CCDW of monolayer H-TaSe$_2$.

The crystal structure of monolayer H-TaSe$_2$ and the optimized atomic displacements in 3×3 CDW are shown in Fig. 1(a) and (b), respectively, where blue arrows indicate the displacement of Ta atoms and circles are Se atoms of upper plane whose out-of-plane displacements are color-coded with reddish (bluish) color for the upward (downward) direction. The CDW is composed of allowed eigenmodes in a 3×3 supercell, which are $Q_1$, $Q_2$, $Q_3$ and $K$ in the first Brillouin zone in Fig. 1(c) where each eigenmode's amplitude is proportional to the area of open circles. Because more than 90% of the total potential energy are from the lowest branches of eigenmodes, we can safely restrict the degrees of freedom to the lowest branch, excluding an eigenmode at $\Gamma$ which corresponds to overall translation. As a result, the low-energy interatomic potential of 3×3 supercell becomes a function of amplitudes $A_i$ and phases $\theta_i$ ($i = 1, 2, 3$ and $K$) of those four eigenmodes. Lowest-branch eigenmodes $Q_1$, $Q_2$, $Q_3$ and $K$ of monolayer H-TMDC is shown in Fig. 1(d), which are obtained in a variational way as described in Appendix D.

All combinations of commensurate eigenmodes satisfying Eq. (2) are listed in Table 1 along with their corresponding energy terms. There are always such combinations in even-order potentials because we can make Eq. (2) to be zero by taking pairs of eigenmodes with the same wavevectors to cancel each other. We call those cases by trivial terms and they do not depend on the phases of eigenmodes. The coefficients of the second, third, and fourth order terms are designated by $B$, $C$, and $D$, respectively, and upper (lower) case implies phase-(in)dependence. $\gamma$ and $\delta$ are reserved for phase constants in third and fourth order terms, respectively. Under the 120º rotation of the crystal, each eigenmode transforms as $Q_1 \to Q_2, Q_2 \to Q_3, Q_3 \to Q_1, K \to K$ with an additional 120° phase advance, and symmetry-equivalent terms share the same coefficients and phase constants. The coefficients and phase constants are extracted from reference density functional calculations whose configurations are constructed by combinations of $Q$ and $K$ eigenmodes with different amplitudes and phases. The details can be found in Appendix E.

The low-energy interatomic potential of commensurate CDW $E_C$ can be explicitly written with all terms in Table 1 and harmonic term $B_Q(A_1{}^2 + A_2{}^2 + A_3{}^2) + B_K A_K{}^2$. We have simplified it by applying the following two relations (see Supplemental Materials) which should be satisfied at the local minima of potentials: 1) $A_1 = A_2 = A_3 = A$, and 2) $(\theta_1, \theta_2, \theta_3) = \left(\theta, \theta, \theta - \frac{2\pi}{3}n\right)$ for $n = 0, 1,$ and 2. Also, we have set $A_K = 0$ and will treat it as a small perturbation later. As a result,

$$E_C = 3B_Q A^2 + 3(D_Q + D_{QQ})A^4 + 3c_Q A^3 f(\theta) \tag{3}$$

where

$$f(\theta) = 3\cos(3\theta + \gamma_Q) + \frac{C_{QQ}}{C_Q}\cos\left(3\theta - \frac{2\pi}{3}n + \gamma_{QQ}\right) \tag{4}$$

. Note the threefold rotational symmetry of potential corresponds to the translational invariance of CDW by a unit cell.

We can expect that for slightly incommensurate CDWs with period $3+\delta$ ($\delta \ll 1$), constituting eigenmodes are composed of ones around the four eigenmodes $Q_1$, $Q_2$, $Q_3$ and $K$. With zeroth-order approximation on higher-order potentials, the potential energy of ICDW $E_{IC}$ can be calculated from $E_C$ by replacing $B_Q$ and $B_K$ with the energies of eigenmodes $\varepsilon(\vec{k})$ calculated from DFPT. For $\varepsilon(\vec{Q}) = B_Q$ to be hold, $\varepsilon(\vec{k})$ is scaled by 0.769. The relative phase transition temperatures scaled by $T_c$ (onset temperature of CDW) is not sensitive to the scaling factor because they are mainly determined by the ratio between $\varepsilon(\vec{Q})$ and $\varepsilon(\vec{Q} + \vec{\delta})$.

### B. Commensurate charge density waves

Equation (3) has multiple local minima, and their structures and energies are reproduced in density functional theory calculations with high accuracies. As we will see later, if temperature effect is incorporated, the energy hierarchy between those local minima can change that commensurate CDWs with different patterns are considered as candidate ground states in the calculation of phase diagram. Figure 2(b) shows $E_n(A, \theta)$ with implicitly optimized $A_K$ and $\theta_K$ for four monolayer H-TMDC's in polar coordinate. The energy spacing of red (green) contour is 1.0 (0.2) meV per unit cell and $E_n(0,0) = 0$ meV. For $n = 0$ and 2, the optimized $A_K = 0$ for all $A$ and $\theta$ while for $n = 1$, there is an energy gain from finite $A_K$. The additional energy gain makes $n = 1$ case a ground state otherwise $n = 2$. Especially for monolayer H-TaSe$_2$, a large energy gain from $A_K$ induces an additional local minimum in $n = 1$ case (two blue indication lines). All local minimum structures are shown in Fig. 2(c) using the same scheme with Fig. 1(b), and the potential energies (dashed lines) are compared with DFT calculations (solid lines) in Fig. 2(a). For monolayer H-TaSe$_2$, the energy gain from $A_K$ decreases that $n = 2$ phase becomes a ground state at finite temperature. Because this transition occurs before the commensurate-incommensurate transition, the first-order commensurate-to-commensurate transition from $n = 1$ to $n = 2$ is expected.

## C. Incommensurate charge density waves

To find the atomic structure and energy of ICDW, we should first limit the number of eigenmodes $n$ for Eq. (1) to have $4n$ variables ($k_x, k_y, A, \theta$ of individual eigenmodes). The difficulty arises from the discontinuous nature $f$ of Eq. (1) on $k_x$ and $k_y$ which greatly deteriorates the convergence of standard optimization method. To make $E_{IC}$ to be differentiable function on $k_x$ and $k_y$, we have set a variational form for ICDW as follows. From the analysis in Appendix F, without any nontrivial relation satisfying Eq. (2), multiple eigenmodes with sufficiently close in *k*-space tend to merge into a single eigenmodes. In other words, if any combinations of eigenmodes are local minima of $E_{IC}$, there should be nontrivial relations Eq. (2) between the eigenmodes with close wavevectors. We used every two eigenmodes around the commensurate wavevectors $Q_1$, $Q_2$, $Q_3$ and $K$, and to satisfy Eq. (2), made their wavevectors to be $\vec{Q} - \vec{\delta}$ and $\vec{Q} + 2\vec{\delta}$. There can be more combinations if we use more eigenmodes, but in our work, the two eigenmodes approximation gives quantitatively good results.

Based on the approximation, we have constructed five variational CDWs as shown in Fig. 3(a). For CCDW, we consider the lowest two cases $n = 1$ (C1) and $n = 2$ (C2) those are consisting of four and three eigenmodes, respectively and the amplitudes and phase of eigenmodes will be optimized according to $E_C$. For ICDWs, we consider the possibility that CDW becomes incommensurate only along specific directions. Three inequivalent cases IC1, IC2 and IC3 which have four, five and six eigenmodes excluding $K$ eigenmode, respectively are possible and their wavevectors, not only amplitudes and phase, are also optimized according to $E_{IC}$. The optimizations are performed using a two-loop conjugate-gradient method where amplitudes and phases of eigenmodes are optimized at inner loop for given wavevectors while wavevectors are optimized at the outer loop.

## D. Temperature-dependent phase diagram of monolayer H-TaSe$_2$

The temperature-dependent phase diagram of monolayer H-TaSe$_2$ has been calculated by comparing the energies of five optimized CDWs. As explained in Sec. **II. C**, the temperature-dependent effective potentials are generated by attaching nearest-neighbor bond-directional springs between Ta and Se atoms whose spring constant is assumed to be linearly proportional to the temperature. The calculated $\varepsilon(\vec{k})$ of lowest-branches are shown in Fig. 3(b) for $T = 0$, and along the high-symmetric lines, $\varepsilon(\vec{k})$ is plotted for $T = 0, 0.5,$ and $1.0$ $T_c$ in Fig. 3(c). At onset temperature of CDW $T_c$, $\varepsilon(\vec{k}) > 0$ for all $\vec{k}$.

Figure 4(a) is the optimized energies of five variational CDWs in monolayer H-TaSe$_2$, and inset shows the energy difference between CDWs. At $T = 0$, $\varepsilon(\vec{k})$ is minimum at $|\vec{k_{min}}| < |\vec{Q}|$ but due to the relation $3\vec{Q} = \vec{G}$, $Q$ eigenmode has additional energy gain from third-order potential. As a result, at $T = 0$, monolayer H-TaSe$_2$ has CCDW as its ground state. Because third-order energy gain gets smaller as amplitudes decreases, in the other three TMDC (H-NbS$_2$, H-TaS$_2$, H-NbSe$_2$) which have relatively weak CDW (see the energy in Fig. 2(a)), ICDW becomes their ground states (not shown here). As temperature increases, the energy gain from $K$ eigenmode becomes smaller that at $T = 0.58\ T_c$, a first-order C1 to C2 phase transition (commensurate-to-commensurate) occurs. At $T = 0.75\ T_c$, IC1 becomes a ground state that commensurate-to-incommensurate phase transition occurs. Interestingly, the transition to incommensurate phase does not simultaneously occur for three symmetry-equivalent directions that CDW remains commensurate to the residing lattice toward other directions (stripe phase). We want to stress that the stripe phase IC1 and IC2 are spontaneously formed at a narrow temperature range between $0.75\ T_c$ and $0.84\ T_c$ without any external effect such as a uniaxial strain. This partially commensurate stripe phase is also observed in the X-ray diffraction measurement of bulk 2H-TaSe$_2$ as a simultaneous occurrence of commensurate and incommensurate peak structure [26, 31]. In actual samples, the three possible commensurate directions are expected to form a domain structure.

The first-order-like transitions are reflected in abrupt change of amplitudes as shown in Fig. 4(b). For IC1 and IC2 $(0.75\ T_c < T < 0.84\ T_c)$, the amplitudes are plotted for both commensurate and incommensurate eigenmodes, and slightly larger for the former ones. $\vec{\delta}$ also shows abrupt jump from zero at $T = 0.75\ T_c$ as in Fig. 4(c). These abrupt transitions are not artifacts of our variational construction of CDW, but the intrinsic nature of commensurate-incommensurate transition where two competing periods coexists in the system. In Appendix G, we explicitly showed the similar abrupt transition occurs in one-dimensional atomic chain without assuming a specific form for ICDW.

Accurately speaking, our results can be applied to a freestanding monolayer TMDC, but we expect they are more similar with bulk 2H-TMDC rather than monolayer TMDC on a substrate where the substrate effect cannot be neglected. The interlayer interaction in bulk 2H-TMDC tends to hinder the out-of-plane displacements of chalcogen atoms, which slightly suppresses the charge density wave. This suppression becomes important when the high pressure is applied to the material. But in ambient pressure, the effect is not so large that we expect various CDW phases in our calculation will also be manifested in bulk 2H-TMDC. In Fig. 4(d), our calculation is compared with the experimental phase diagram of bulk 2H-TaSe$_2$ using X-ray diffraction [31]. The observed hysteresis in the experiment is presumably due to the first-order transition between IC structures.

We cannot identify the presence of two distinct CCDW phases in the X-ray diffraction due to the absence of low-temperature data. In actual defective samples, the small energy difference between C1 and C2 phases (at most 0.04 meV/unitcell from the inset if Fig. 4(a)) allows the coexistence of both phases, and the relative abundance of them can be measured in local probe technics to verify our prediction. The presence of three distinct ICDW is partly proved in the experiment, because IC1 and IC2 are not distinguished in the domain-averaged information. In spite that the calculation can quantitatively predict transition temperatures and various CDW phases, $|\vec{\delta}|$ in Fig. 4(c) is still larger than the experimental value by a factor 2-3 (not shown here). From the total energy differences between various phases, one can estimate the require accuracy of interatomic potential is quite stringent. Especially, $\vec{\delta}$ strongly depends on the long-range harmonic potential which is the crudest part of our approximation.

In summary, we have developed an interatomic potential method enabling the practical calculation of CDW phase diagrams by eigenmode representation. Our calculation quantitatively reproduced the relative temperature of multiple CDW phase transition and the emergence of stripe phases without any adjustable parameters. The main source of error in our calculation is from the use of long-range harmonic potential obtained from perturbative calculation. The results that complex behaviors of charge density wave are originated from the relatively simple structure of the interatomic potential, will shed light on the current understanding of CDW and its control using pressure or metal intercalation.

**ACKNOWLEDGEMENT**

We thank Hosub Jin and Young-Woo Son for helpful discussions. C. P. was supported by the New generation research program (CG079701) at Korea Institute for Advanced Study (KIAS). Computations were supported by Center for Advanced Computation of KIAS.

**APPENDIX A: DERIVATION OF EQUATION 1**

Equation (1) can be derived by a change of variables from atomic displacement to eigenmodes. Specifically, consider the third-order energy $E_3$ (similar for higher-order energies) expressed as

$$E_3(\{r_\alpha(\vec{R})\}) = \frac{1}{N} \sum_{\{\vec{R}\},\{\alpha\}} C_{\overrightarrow{R_1},\overrightarrow{R_2},\{\alpha\}} r_{\alpha_0}(\overrightarrow{R_0}) r_{\alpha_1}(\overrightarrow{R_0} + \overrightarrow{R_1}) r_{\alpha_2}(\overrightarrow{R_0} + \overrightarrow{R_2}) \tag{A1}$$

where $N$ is a number of unitcell in a sample, $\vec{R}$ the lattice vectors of an unitcell, $\alpha$ a condensed index of Cartesian component and basis atom in a unitcell, $r_\alpha(\vec{R})$ a Cartesian component of atomic displacement in an unitcell at $\vec{R}$, and $C$ is the force constant. $\{\vec{R}\}$ and $\{\alpha\}$ mean $\vec{R_0}, \vec{R_1}, \vec{R_2}$ and $\alpha_0, \alpha_1, \alpha_2$, respectively. Because $C$ respects the translational symmetry of a crystal, it does not depend on $\vec{R_0}$. When $n$ eigenmodes of amplitudes $A_i$, phase $\theta_i$, wavevector $\vec{k_i}$ and (complex) polarization vector $\vec{p_i}$ whose component $\alpha$ is written as $p_{i,\alpha}e^{i\theta_\alpha}$ are given, atomic displacements can be written as

$$r_\alpha(\vec{R}) = \sum_{i=1}^{n} A_i\, p_{i,\alpha} Re[e^{i(\vec{k_i}\cdot\vec{R}+\theta_i+\theta_\alpha)}] \tag{A2}$$

. Inserting Eq. (A2) into Eq. (A1), $E_3$ becomes

$$E_3(A_1,\cdots,A_n,\theta_1,\cdots,\theta_n) = \sum_{\vec{R_1},\vec{R_2},\{\alpha\}} C_{\vec{R_1},\vec{R_2},\{\alpha\}} \sum_{i,j,k=1}^{n} A_i A_j A_k\, p_{i,\alpha_0} p_{j,\alpha_1} p_{k,\alpha_2}\, g_{ijk} \tag{A3}$$

where

$$g_{ijk} = \frac{1}{2^3} \sum_{s_0,s_1,s_2=0}^{1} \left( e^{i\phi(s_0,s_1,s_2)} \times \frac{1}{N}\sum_{\vec{R_0}} e^{i\{(2s_0-1)\vec{k_i}+(2s_1-1)\vec{k_j}+(2s_2-1)\vec{k_k}\}\cdot\vec{R_0}} \right) \tag{A4}$$

. Here,

$$\phi(s_0,s_1,s_2) = (2s_0-1)(\theta_i+\theta_{\alpha_0}) + (2s_1-1)(\theta_j+\theta_{\alpha_1}+\vec{k_j}\cdot\vec{R_1}) + \\ (2s_2-1)(\theta_k+\theta_{\alpha_2}+\vec{k_k}\cdot\vec{R_2}) \tag{A5}$$

. The term in the parenthesis of Eq. (A4) become $2\cos\phi(s_0,s_1,s_2)$ for $(2s_0-1)\vec{k_i}+(2s_1-1)\vec{k_j}+(2s_2-1)\vec{k_k} = n\vec{G}$ ($n$ is an integer and $\vec{G}$ is a reciprocal lattice vector), or 0 for else. By comparing Eq. (3) with Eq. (A3),

$$c_{ijk} = 2\sum_{\vec{R_1},\vec{R_2},\{\alpha\}} C_{\vec{R_1},\vec{R_2},\{\alpha\}} p_{i,\alpha_0} p_{j,\alpha_1} p_{k,\alpha_2} \tag{A6}$$

$$\gamma_{ijk} = (2s_0 - 1)\theta_{\alpha_0} + (2s_1 - 1)(\theta_{\alpha_1} + \vec{k_j} \cdot \vec{R_1}) + (2s_2 - 1)(\theta_{\alpha_2} + \vec{k_k} \cdot \vec{R_2}) \quad (A7)$$

.

**APPENDIX B: ANHARMONIC COUPLING COEFFICIENTS BETWEEN EIGENMODES WITH SIMILAR WAVEVECTORS**

Consider a commensurate eigenmode $\rho$ with amplitudes $A$, phase $\theta$, wavevector $\vec{k} = \frac{1}{3}\vec{G}$ and (complex) polarization vector $\vec{p}$ whose component $\alpha$ is written as $p_\alpha e^{i\theta_\alpha}$ where $\alpha$ is a condensed index of Cartesian components and atoms in a unitcell. We will concisely write $\rho = (A, \theta, \vec{k}, \vec{p})$. The third and fourth order energy $E_3$ and $E_4$ of $\rho$ becomes

$$\begin{aligned} E_3 &= A^3 \sum_{\overrightarrow{R_1},\overrightarrow{R_2},\{\alpha\}} \frac{1}{4} C_{\overrightarrow{R_1},\overrightarrow{R_2},\{\alpha\}} \, p_{\alpha_0} p_{\alpha_1} p_{\alpha_2} Re\left[e^{i(3\theta + \vec{k}\cdot\vec{R_1} + \vec{k}\cdot\vec{R_2} + \theta_{\alpha_0} + \theta_{\alpha_1} + \theta_{\alpha_2})}\right] \\ &\equiv cA^3 \cos(3\theta + \gamma) \end{aligned} \quad (B1)$$

$$\begin{aligned} E_4 = A^4 \sum_{\overrightarrow{R_1},\overrightarrow{R_2},\overrightarrow{R_3},\{\alpha\}} \frac{1}{8} C_{\overrightarrow{R_1},\overrightarrow{R_2},\overrightarrow{R_3}\{\alpha\}} \, p_{\alpha_0} p_{\alpha_1} p_{\alpha_2} p_{\alpha_3} \\ \times \left[ Re\left[e^{i(\vec{k}\cdot\vec{R_1} - \vec{k}\cdot\vec{R_2} - \vec{k}\cdot\vec{R_3} + \theta_{\alpha_0} + \theta_{\alpha_1} - \theta_{\alpha_2} - \theta_{\alpha_3})}\right] \right. \\ + Re\left[e^{i(-\vec{k}\cdot\vec{R_1} + \vec{k}\cdot\vec{R_2} - \vec{k}\cdot\vec{R_3} + \theta_{\alpha_0} - \theta_{\alpha_1} + \theta_{\alpha_2} - \theta_{\alpha_3})}\right] \\ \left. + Re\left[e^{i(-\vec{k}\cdot\vec{R_1} - \vec{k}\cdot\vec{R_2} + \vec{k}\cdot\vec{R_3} + \theta_{\alpha_0} - \theta_{\alpha_1} - \theta_{\alpha_2} + \theta_{\alpha_3})}\right] \right] \equiv dA^4 \end{aligned} \quad (B2)$$

Now consider two incommensurate modulations with $\rho_1 = (A_1, \theta_1, \vec{k} - \vec{\delta}, \vec{p} + \overrightarrow{p_1})$ and $\rho_2 = (A_2, \theta_2, \vec{k} + 2\vec{\delta}, \vec{p} + \overrightarrow{p_2})$ where $\vec{\delta}$, $\overrightarrow{p_1}$ and $\overrightarrow{p_2}$ are small. By applying a zeroth order approximation to $\vec{p}$ ($\overrightarrow{p_1} = \overrightarrow{p_2} = 0$), the third order energy $E_3'$ of $\rho_1 + \rho_2$ are as the following.

$$\begin{aligned} E_3' = A_1^2 A_2 \sum_{\overrightarrow{R_1},\overrightarrow{R_2},\{\alpha\}} \frac{1}{4} C_{\overrightarrow{R_1},\overrightarrow{R_2},\{\alpha\}} \, p_{\alpha_0} p_{\alpha_1} p_{\alpha_2} \\ \times \left[ Re\left[e^{i(2\theta_1 + \theta_2 + \vec{k}\cdot\vec{R_1} + \vec{k}\cdot\vec{R_2} + \theta_{\alpha_0} + \theta_{\alpha_1} + \theta_{\alpha_2})} e^{i(-\vec{\delta}\cdot\vec{R_1} + 2\vec{\delta}\cdot\vec{R_2})}\right] \right. \\ + Re\left[e^{i(2\theta_1 + \theta_2 + \vec{k}\cdot\vec{R_1} + \vec{k}\cdot\vec{R_2} + \theta_{\alpha_0} + \theta_{\alpha_1} + \theta_{\alpha_2})} e^{i(2\vec{\delta}\cdot\vec{R_1} - \vec{\delta}\cdot\vec{R_2})}\right] \\ \left. + Re\left[e^{i(2\theta_1 + \theta_2 + \vec{k}\cdot\vec{R_1} + \vec{k}\cdot\vec{R_2} + \theta_{\alpha_0} + \theta_{\alpha_1} + \theta_{\alpha_2})} e^{i(-\vec{\delta}\cdot\vec{R_1} - \vec{\delta}\cdot\vec{R_2})}\right] \right] \end{aligned} \quad (B3)$$

By further approximating $\vec{\delta} = 0$, $E_3' = \mathbf{3}cA_1{}^2A_2\cos(2\theta_1 + \theta_2 + \gamma)$ holds. Note the factor **3** is multiplied compared with $E_3 = cA^3\cos(3\theta + \gamma)$. A similar calculation gives factor **4** for the fourth order energy $E_4' = dA_1{}^4 + dA_2{}^4 + \mathbf{4}dA_1{}^2A_2{}^2$.

## APPENDIX C: TEMPERATURE-DEPENDENT EFFECTIVE POTENTIAL FOR AN ATOMIC CHAIN

At finite temperature $T$, fluctuating atoms can be described by their probability distribution functions $\sigma_T(r)$. If we exert forces on $\sigma_T(r)$'s, the centers of them deviate from their equilibrium position. The effective potential energy at $T$ is defined as the energy required to keep the centers of $\sigma_T(r)$ at the deviated positions. We made the following two approximations for $\sigma_T(r)$: (1) $\sigma_T(r)$ respects all the symmetries of the reference configuration, and (2) $\sigma_T(r)$ rigidly shifts as the center displaces. Neglecting the correlations between $\sigma_T(r)$'s (mean-field approximation), the potential energy of the displaced system at $T$ is calculated with integrations between $\sigma_T(r)$'s.

For a specific example, consider a one-dimensional atomic chain with $i$-th neighbor spring constant $k_i$ ($i =$ 1, 2, 3) and nearest-neighbor third- (fourth-) order force constant $c$ ($d$) depicting the short-range anharmonicity. For this chain, the potential energy per atom $E$ can be written as

$$E = \frac{1}{N}\sum_{i}^{N}\left[\sum_{j=1}^{3}k_j(\Delta x_i - \Delta x_{i-j})^2 + c(\Delta x_i - \Delta x_{i-1})^3 + d(\Delta x_i - \Delta x_{i-1})^4\right] \quad (C1)$$

where $\Delta x_i$ is a displacement of $i$-th atom from its reference configuration. The fourth-order potential energy at finite temperature is calculated as

$$\begin{aligned}E_4(T) &= \frac{1}{N}\sum_{i}^{N} d\int dX_i dX_{i-1}(X_i - X_{i-1})^4\sigma_T(X_i - x_i)\sigma_T(X_{i-1} - x_{i-1})\\ &= \frac{1}{N}\sum_{i}^{N}[d(x_i - x_{i-1})^4 + 6d(x_i - x_{i-1})^2 I_2 + I_4]\end{aligned} \quad (C2)$$

where

$$I_n = \int dx_1 dx_2 (x_1 - x_2)^n \sigma_T(x_1) \sigma_T(x_2). \tag{C3}$$

Note $I_n = 0$ for odd $n$ because $\sigma_T(x) = \sigma_T(-x)$ and $I_n \geq 0$ for even $n$. Comparing $E_4(T)$ with $E_4(T=0) = \frac{1}{N}\sum_i^N d(x_i - x_{i-1})^4$, the fourth-order term results in the effective second-order term $6d(x_i - x_{i-1})^2 I_2$ which can be interpreted as an additionally attached nearest-neighbor spring. $I_4$ is not relevant to effective force constant but just redefines the reference of zero energy. Similarly, one can check that $E_2(T) = E_2(T=0)$ and $E_3(T) = E_3(T=0)$. If we further approximate $\sigma_T(x)$ with a gaussian as $\sigma_T(x) = \frac{1}{N_0}\exp\left(-\frac{\alpha x^2}{k_B T}\right)$ where $k_B$, $N_0$, and $\alpha$ are Boltzmann constant, normalization factor, and temperature-independent constant determining the critical onset temperature of CDW, respectively, $I_2$ becomes linearly proportional to temperature and the linearity holds for higher-dimensional distributions.

## APPENDIX D: VARIATIONAL CALCULATION OF THE LOWEST-BRANCH EIGENMODES OF MONOLAYER H-TMDC

Thanks to the symmetry of monolayer H-TMDC, polarization vectors $\vec{p}$ of the lowest-branch eigenmodes can be written as

$$\vec{p} = \left(M_x, M_y, M_z, C_x^{up}, C_y^{up}, C_z^{up}, C_x^{dn}, C_y^{dn}, C_z^{dn}\right)$$
$$= \begin{cases} N_Q\left(V_x, V_y, 0, V_x r_{xy}^Q e^{i\theta_{xy}}, V_y r_{xy}^Q e^{i\theta_{xy}}, r_z^Q e^{i\theta_z}, V_x r_{xy}^Q e^{i\theta_{xy}}, V_y r_{xy}^Q e^{i\theta_{xy}}, -r_z^Q e^{i\theta_z}\right), & \text{for } Q \\ N_K\left(V_x, V_y, 0, -V_x r_{xy}^K, -V_y r_{xy}^K, 0, -V_x r_{xy}^K, -V_y r_{xy}^K, 0\right), & \text{for } K \end{cases} \tag{D1}$$

where $M$ and $C^{up(dn)}$ stand for metal and upper (lower) chalcogen atoms in a unitcell, $N$ is a normalization factor and $(V_x, V_y) = \left(\frac{\sqrt{3}}{2}, \frac{1}{2}\right), \left(\frac{\sqrt{2}}{2}, \frac{\sqrt{2}}{2} i\right)$ for $Q_1$ and $K$, respectively. The atomic displacement of the eigenmode with amplitude $A$ and phase $\theta$ becomes $Re\left[A e^{i\theta} e^{i\vec{k}\cdot\vec{R}} \vec{p}\right]$ where $\vec{R}$ is a lattice vector indexing a unitcell position. For fixed amplitudes and phases of $\vec{p}$, materials-dependent parameters $r_{xy}^Q, r_z^Q, \theta_{xy}, \theta_z, r_{xy}^K$ are obtained in 3×3 supercell calculation minimizing the total energy. For an eigenmode $Q_1$ ($K$), 12 (9) combinations of $A$ and $\theta$ are used for parameters as shown in points in Fig. 5, and their average values as listed in Table 2.

## APPENDIX E: FITTING METHOD FOR COEFFICIENTS AND PHASE CONSTANTS OF INTERATOMIC POTENTIAL

The coefficients and phase constants of 3×3 1H-TMDC potential are fitted to the reference density functional theory calculations. Specifically, three independent least-square fittings are performed. In Table 4, the first three configurations are only composed of $Q$ eigenmodes with 0.03 Å $\leq A \leq$ 0.10 Å and 0 $\leq \theta \leq$ 100° and the fourth ones are composed of $K$ eigenmodes with 0.02 Å $\leq A \leq$ 0.08 Å and 0° $\leq \theta \leq$ 100°. The formers are simultaneously fitted to the corresponding combinations of $E_{Q1}$, $E_{Q2}$ and $E_{Q3}$ in Table 4 and the left five panels in Fig. 8 shows the reference energy (points) and fitted functions (lines) of them. Because we are interested in the low-energy potential, configurations with larger energy than 2 meV/unitcell are excluded from the fitting procedure. The right panels in Fig. 6 are similar plots for $K$ eigenmodes.

The coupling coefficients between $Q$ and $K$ are obtained from the fifth configurations in Table 4. The fitting procedures are outlined in Fig. 7 for the case of monolayer H-TaSe$_2$. The reference configurations are constructed with superpositions of where three $Q$ eigenmodes and $K$ eigenmodes. We first fix the amplitudes and phases of three $Q$ eigenmodes as $A_Q = $ 0.04 Å, and $\pi/2, \pi/2$ and $-\pi/6$ ($n = 1$ case), respectively. By varying the amplitudes and phases of $K$ eigenmodes as -0.04 Å $\leq A_K \leq$ 0.04 Å and phases 0° $\leq \theta_K \leq$ 330°, $E_{K,n=1}(A_Q, A_K, \theta)$ (blue lines in Fig. 7(a)) is fitted to the reference potential energy (red points in Fig. 7(a)). The linear and quadratic coefficients of polynomial in Fig. 7(a) are again fitted to $E_{A1}(A_Q, \theta_K)$ (Fig. 9(b)) and $E_{A2}(A_Q, \theta_K)$ (Fig. 9(c)), which contain parameters $c_{QK}$, $d_{QK1}$, $D_{QK}$, and $d_{QK2}(\approx 0)$. To obtain the individual parameters from functions such as $3c_{QK}A_Q^2 + 3d_{QK1}A_Q^3$ and $3D_{QK}A_Q^2$, we repeated the whole procedure for different $A_Q$. The least square fitting results for the parameters are plotted in Fig. 7(d) for four monolayer H-TMDC's.

## APPENDIX F: STABILITY CRITERION OF MULTIPLE EIGENMODES

Consider $N$ eigenmodes with the same energy $\varepsilon < 0$. If there is no nontrivial relation between their wavevectors satisfying Eq. (2), the potential energy $E$ of the eigenmodes up to fourth order can be written as

$$E = \varepsilon \sum_{i=1}^{N} A_i^2 + D_2 \sum_{i \neq j} A_i^2 A_j^2 + D_1 \sum_{i=1}^{N} A_i^4 \tag{F1}$$

where $A_i$ is the amplitude of $i$-th eigenmode, $D_0 > 0$ and $D_1 > 0$ are fourth-order coefficients. Equation (F1) can describe either the potential energy of symmetry-equivalent eigenmodes or that of multiple eigenmodes with almost the same wavevectors. The ground state of Eq. (F1) is

$$\begin{cases} A_1^2 = -\dfrac{\varepsilon}{2D_1}, A_2, A_3, \cdots, A_N = 0, & 2D_1 < D_2 \\ A_1^2 = A_2^2 = \cdots = A_N^2 = -\dfrac{\varepsilon}{2D_1 + (N-1)D_2}, & 2D_1 \geq D_2 \end{cases} \quad (F2)$$

. The first implication of Eq. (F2) is that the rotational symmetry of CDW can be spontaneous broken. This suggests a criterion for one- and two-dimensional CDW formation in crystals. Interestingly, $D_2 = 2D_1$ and $D_2 = 4D_1$ for a simple triangular and square lattice with solely nearest-neighbor bond-directional fourth-order repulsive interactions. When other effects such as anharmonic interaction between basis atoms and non-bond-directional anharmonicity are involved, the value can change, but it can explain why many CDWs in square lattices become one-dimensional [32-34], but only a few [35] in triangular lattices.

The second implication of Eq. (F2) is that if two or more eigenmodes are sufficiently close in $k$-space ($D_2 \approx 4D_1$, see Appendix B), they are unstable to the merging into a single eigenmode unless any $A$-linear terms from nontrivial relation stabilize the multiple eigenmodes. In other words, if any CDW is composed of eigenmodes with sufficiently close wavevectors, the wavevectors satisfy the relation Eq. (2).

**APPENDIX G: COMMENSURATE-INCOMMENSURATE PHASE TRANSITION IN ATOMIC CHAIN**

how the lattice anharmonicity becomes the origin of the commensurate period and what is the structure of an ICDW.

We consider one-dimensional atomic chain whose interatomic potential is defined as Eq. (C1). Figure 8(a) shows the atomic chain and we have set $k_1 = 0.3$ Ry/bohr², $k_2 = -0.4$ Ry/bohr², $k_3 = 0.5$ Ry/bohr², $c = -100$ eV/Å³ and $d = 1000$ eV/Å⁴. The red line in Fig. 8(b) is the dispersion relation $\varepsilon(k) = 2\sum_{i=1}^{3} k_i \sin^2 \frac{ik}{2}$ of the chain. The chain is unstable to periodic distortions with wavevectors $k$ around $\frac{2\pi}{3a}$ ($a$ is the interatomic distance of the reference configuration). The ground state of the chain is found among tens of local minimum structures which are prepared by the following procedures:

random initial configuration generation, pre-optimization by a nonlinear conjugate gradient method, low-frequency Fourier filtering, and re-optimization. We gradually increased the size of supercell and plotted their optimized energies as red dots in Fig. 8(d) according to their largest Fourier component of displacements. The period of the ground state is 3, which slightly deviates from the minimum of $\varepsilon(k)$. Decomposing the potential energy into contributions from eigenmodes, which are equivalent to Fourier components in this case, reveals the origin of commensurate ground state. By inserting the eigenmode with the largest contribution $\Delta x_i = A \cos(ika + \theta)$ into Eq. (C1), the potential energy becomes

$$E(A, \theta, k) = \varepsilon(k)A^2 + 6d \sin^4 \frac{ka}{2} A^4 + 2c \sin^3 \frac{ka}{2} \sin(3\theta - \frac{3ka}{2}) A^3 \delta_{\frac{2\pi}{3a}, k} \quad (G1)$$

for $k > \frac{\pi}{2a}$ where $\delta$ is a Kronecker delta function. By minimizing $E$ with respect to $A$ and $\theta$ for a given $k$, the potential energy $E(k)$ is plotted in Fig. 8(d) with a black solid line and a single discrete point at $k = \frac{2\pi}{3a}$. The last term in Eq. (G1) provides a discontinuous energy gain for the commensurate period 3. Even if this discontinuity smooths out in actual optimized energies, the tendency favoring commensurate period persists.

The incommensurate solutions are worth noting because they can become ground states for other parameter range or local stable structures in the presence of immobile defects. Figure 8(c) is the displacement pattern of a selected incommensurate structure indicated by an arrow in Fig. 8(d) and the structure can be understood as a discommensuration [15] where local commensurate regions are separated by soliton-like domain walls. The width of a soliton is determined by how much the system favors commensurate structures and for our case, the width is sufficiently small that inter-soliton interaction can be safely neglected. Because the soliton density of incommensurate structure with $k = \left(\frac{2}{3} - \delta\right)\frac{\pi}{a}$ is $\frac{\delta}{2a}$, neglecting inter-soliton interactions, the potential energy per atom becomes $E(\delta) = E_0 + \frac{\delta}{2a} E_s$ where $E_0$ is the energy of period 3 structure and $E_s$ is the energy of a single soliton. This explains the linear $E(k)$ relation around $k = \frac{2\pi}{3a}$.

First-order-like commensurate-incommensurate phase transition occurs as the temperature increases. The temperature effect is incorporated by increasing the nearest-neighbor spring constant by $\Delta k_1$. For $\Delta k_1 = 0.18$ Ry/bohr$^2$, all eigenmodes have positive energies and this defines the onset temperature $T_c$ of a lattice modulation. Using the linear relation between $\Delta k_1$ and temperature, the temperature will be defined as $T = \frac{\Delta k_1}{0.18} T_c$. Figure 8(b) shows the change of $\frac{1}{2} m\omega^2(k)$ on the temperature. In Fig. 9(a), the ground states at given temperatures are calculated similarly with $T = 0$ case. As temperature increases from 0, the linear dispersion relation around $k = \frac{2\pi}{3a}$ starts to break due to the interaction between broadened solitons and at $T = 0.172\, T_c$, the period locked-in $k = \frac{2\pi}{3a}$ abruptly changes to $k = \left(\frac{2}{3} - \delta\right)\frac{\pi}{a}$ for $\delta = 0.04$. Around the transition, the energies of periodic modulations with $k$ between $\left(\frac{2}{3} - \delta\right)\frac{\pi}{a}$ and $\frac{2\pi}{3a}$ are almost similar that even weak defects can easily destroy the global ordering of the chain.

The ground state structures at finite temperatures are shown in Fig. 9(c) as Fourier transformations of the atomic displacements. The incommensurate ground states have only a few eigenmodes, and the main eigenmode with $k_m = \left(\frac{2}{3} - \delta\right)\frac{\pi}{a}$ always accompanies the satellite eigenmode with $k_s = \left(\frac{2}{3} + 2\delta\right)\frac{\pi}{a}$ whose presence lowers the potential energy as the follows. The relation $2k_m + k_s = \frac{2\pi}{a}$ results in energy term proportional to $A_m^2 A_s \sin(2\theta_m + \theta_s)$ where $A_m(A_s)$ and $\theta_m(\theta_s)$ stand for the amplitude and phase of main (satellite) eigenmodes, respectively. For an appropriate $\theta_s$, finite $A_s$ can lower the energy of the structure because it is the lowest order term in $A_s$ among all energy terms from the two eigenmodes. The main and satellite eigenmode cover more than 95% of the total potential energy. The amplitudes of main eigenmode undergo abrupt decrease at commensurate-incommensurate transition as in Fig. 9(c), reflecting the first-order nature of the transition.

At slightly below $T_c$, the period of a lattice modulation is close to the minimum of $\varepsilon(k, T = T_c)$ due to the reduced anharmonic effect for a small amplitude. This is the reason why the analysis based on the shape of Fermi surface, which misses the role of the anharmonicity, seems to be valid for inferring the period of an incipient or weak CDW in a metal. Because there is no reason for this period favoring an integer value except accidental cases, most incipient periodic lattice modulations tend to be incommensurate to the residing lattice.

| Relation | Energy term |
|---|---|
| Trivial (phase-independent) | |
| $2Q_i - 2Q_i$ | $D_Q A_i^4$ |
| $Q_i - Q_i + Q_j - Q_j$ | $D_{QQ} A_i^2 A_j^2$ |
| $2K - 2K$ | $D_K A_K^4$ |
| $Q_i - Q_i + K - K$ | $D_{QK} A_i^2 A_K^2$ |
| Nontrivial (phase-dependent) | |
| $3Q_i$ | $c_Q A_i^3 \cos(3\theta_i + \gamma_Q)$ |
| $Q_1 + Q_2 + Q_3$ | $c_{QQ} A_1 A_2 A_3 \cos(\theta_1 + \theta_2 + \theta_3 + \gamma_{QQ})$ |
| $3K$ | $c_K A_K^3 \cos(3\theta_K + \gamma_K)$ |
| $K - Q_1 + Q_2$ | $c_{QK} A_K A_1 A_2 \cos(\theta_K - \theta_1 + \theta_2 + \gamma_{QK})$ |
| $K - Q_2 + Q_3$ | $c_{QK} A_K A_2 A_3 \cos(\theta_K - \theta_2 + \theta_3 + \gamma_{QK} + 2\pi/3)$ |
| $K - Q_3 + Q_1$ | $c_{QK} A_K A_3 A_1 \cos(\theta_K - \theta_3 + \theta_1 + \gamma_{QK} + 4\pi/3)$ |
| $2Q_1 - Q_2 - Q_3$ | $d_{QQ} A_1^2 A_2 A_3 \cos(2\theta_1 - \theta_2 - \theta_3 + \delta_{QQ})$ |
| $2Q_2 - Q_3 - Q_1$ | $d_{QQ} A_2^2 A_3 A_1 \cos(2\theta_2 - \theta_3 - \theta_1 + \delta_{QQ})$ |
| $2Q_3 - Q_1 - Q_2$ | $d_{QQ} A_3^2 A_1 A_2 \cos(2\theta_3 - \theta_1 - \theta_2 + \delta_{QQ})$ |
| $K - 2Q_1 - Q_3$ | $d_{QK1} A_K A_1^2 A_3 \cos(\theta_K - 2\theta_1 - \theta_3 + \delta_{QK1})$ |
| $K - 2Q_2 - Q_1$ | $d_{QK1} A_K A_2^2 A_1 \cos(\theta_K - 2\theta_2 - \theta_1 + \delta_{QK1} + 2\pi/3)$ |
| $K - 2Q_3 - Q_2$ | $d_{QK1} A_K A_3^2 A_2 \cos(\theta_K - 2\theta_3 - \theta_2 + \delta_{QK1} + 4\pi/3)$ |
| $2K + Q_1 - Q_2$ | $d_{QK2} A_K^2 A_1 A_2 \cos(2\theta_K + \theta_1 - \theta_2 + \delta_{QK2})$ |
| $2K + Q_2 - Q_3$ | $d_{QK2} A_K^2 A_2 A_3 \cos(2\theta_K + \theta_2 - \theta_3 + \delta_{QK2} - 2\pi/3)$ |
| $2K + Q_3 - Q_1$ | $d_{QK2} A_K^2 A_3 A_1 \cos(2\theta_K + \theta_3 - \theta_1 + \delta_{QK2} - 4\pi/3)$ |

Table 1. Three and four combinations of $\vec{Q_i}$ and $\vec{K}$ satisfying Eq. (2). Corresponding energy terms are listed on the right column.

|  | NbS$_2$ | NbSe$_2$ | TaS$_2$ | TaSe$_2$ |
| --- | --- | --- | --- | --- |
| $a$ (Å) | 3.358 | 3.484 | 3.336 | 3.468 |
| $\Delta h$ (Å) | 1.567 | 1.684 | 1.562 | 1.673 |
| $r_{xy}^Q$ | 0.30 | 0.23 | 0.29 | 0.23 |
| $r_z^Q$ | 0.35 | 0.46 | 0.45 | 0.44 |
| $r_{xy}^K$ | 0.93 | 0.66 | 0.58 | 0.41 |
| $\theta_{xy}$ (°) | 125 | 112 | 126 | 116 |
| $\theta_z$ (°) | -25 | -25 | -25 | -25 |

Table 2. Optimized lattice constants ($a$) and height differences between transition metal and chalcogen atom ($\Delta h$) in unitcell calculations. Configuration-averaged polarization parameters of Q ($r_{xy}^Q$, $r_z^Q$, $\theta_{xy}$, $\theta_z$) and K ($r_{xy}^K$) eigenmodes are presented for four monolayer H-TMDC's.

|  | $NbS_2$ | $TaS_2$ | $NbSe_2$ | $TaSe_2$ |
|---|---|---|---|---|
| $B_Q$ | -9.0 | -10.3 | -6.2 | -8.4 |
| $B_K$ | 28.3 | 10.1 | 16.1 | 4.4 |
| $D_Q$ | 20.9 | 20.8 | 8.2 | 7.5 |
| $D_K$ | 0.6 | 3.8 | 0.0 | 0.2 |
| $D_{QQ}$ | 18.0 | 18.5 | 8.3 | 6.8 |
| $D_{QK}$ | 41.9 | 22.9 | 21.3 | 9.9 |
| $c_Q, \gamma_Q$ | 2.1, 6° | 3.4, 22° | 1.3, 305° | 1.3, 355° |
| $c_K, \gamma_K$ | 3.6, 270° | 1.9, 270° | 0.0, 270° | 1.4, 270° |
| $c_{QQ}, \gamma_{QQ}$ | 6.4, 197° | 5.7, 166° | 3.8, 183° | 3.7, 199° |
| $c_{QK}, \gamma_{QK}$ | 3.7, 90° | 7.3, 90° | 4.9, 90° | 3.7, 90° |
| $d_{QQ}, \delta_{QQ}$ | 1.5, 17° | 0.5, 345° | 1.2, 15° | 0.4, 19° |
| $d_{QK1}, \delta_{QK1}$ | 15.5, 240° | 10.4, 240° | 10.4, 240° | 13.7, 240° |
| $d_{QK2}, \delta_{QK2}$ | ~0 | ~0 | ~0 | ~0 |

Table 3. Calculated potential coefficients of monolayer H-TMDC's. Unit is meV/(0.1 Å)$^n$ where $n = 2$, 3 and 4 for $B$, $c$ and $D$, respectively.

| Configuration | Fitting function |
|---|---|
| $A_1 = A$ <br> $\theta_1 = \theta$ | $E_{Q1}(A, \theta)$ |
| $A_1 = A_2 = A$ <br> $\theta_1 = \theta_2 = \theta$ | $2E_{Q1}(A, \theta) + E_{Q2}(A)$ |
| $A_1 = A_2 = A_3 = A$ <br> $\theta_1 = \theta_2 = \theta, \theta_3 = \theta - 2\pi n/3$ | $3E_{Q1}(A, \theta) + 3E_{Q2}(A) + E_{Q3}(A, \theta, n)$ |
| $A_K = A$ <br> $\theta_K = \theta$ | $E_{K1}(A, \theta)$ |
| $A_1 = A_2 = A_3 = \pm A_Q, A_K = A$ <br> $\theta_1 = \theta_2 = \pi/2, \theta_3 = -\pi/6, \theta_K = \theta$ | $E_{K,n=1}(A_Q, A, \theta)$ |
| Function forms ||

$E_{Q1}(A, \theta) = B_Q A^2 + c_Q r^3 \cos(3\theta + \gamma_Q) + D_Q A^4$

$E_{Q2}(A) = D_{QQ} A^4$

$E_{Q3}(A, \theta, n) = c_{QQ} A^3 \cos(3\theta + \gamma_{QQ} - 2\pi n/3) + d_{QQ} A^4 \cos(2\pi n/3 + \delta_{QQ})$

$E_{K1}(A, \theta) = B_K A^2 + c_K r^3 \cos(3\theta + \pi/2) + D_K A^4$

$E_{K,n=1}(A_Q, A, \theta_K) = E_0 + A E_{A1}(A_Q, \theta_K) + A^2 E_{A2}(A_Q, \theta_K)$

$E_{A1}(A_Q, \theta_K) = 3c_{QK} A_Q^2 \cos(\theta_K + \gamma_{QK}) + 3d_{QK1} A_Q^3 \cos(\theta_K - 5\pi/6 + \delta_{QK1})$

$E_{A2}(A_Q, \theta_K) = 3D_{QK} A_Q^2 + d_{QK2} A_Q^2 \cos(2\theta_K + \delta_{QK2})$

Table 4. (Upper left column) Reference configurations constructed by combinations of $Q$ and $K$ eigenmodes. Amplitudes and phases of $Q(K)$ eigenmodes are $A_i(A_K)$ and $\theta_i(\theta_K)$, respectively. (Upper right column) Fitting functions used for corresponding configurations. Explicit function forms with coefficients are shown in the bottom row.

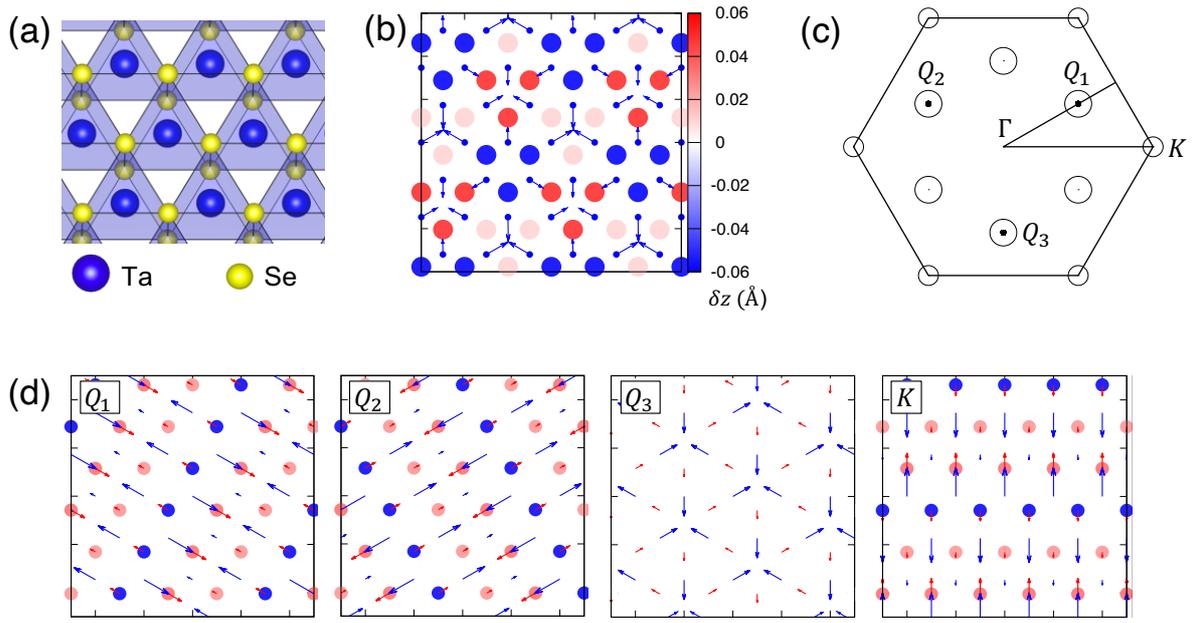

FIG. 1. (a) Crystal structure of monolayer 1H-TaSe$_2$ and (b) calculated displacement vectors of atoms in the 3×3 CDW structure where blue arrows indicate the displacement of Ta atoms and circles are Se atoms of upper plane whose out-of-plane displacements are color-coded with reddish (bluish) color for the upward (downward) direction. (c) Displacements vectors in (b) are projected on the eigenmodes of monolayer H-TaSe$_2$ and plotted in the first Brillouin zone by open circles. The areas of the circles are proportional to the amplitudes and the positions of the circles denote wavevectors of eigenmodes.

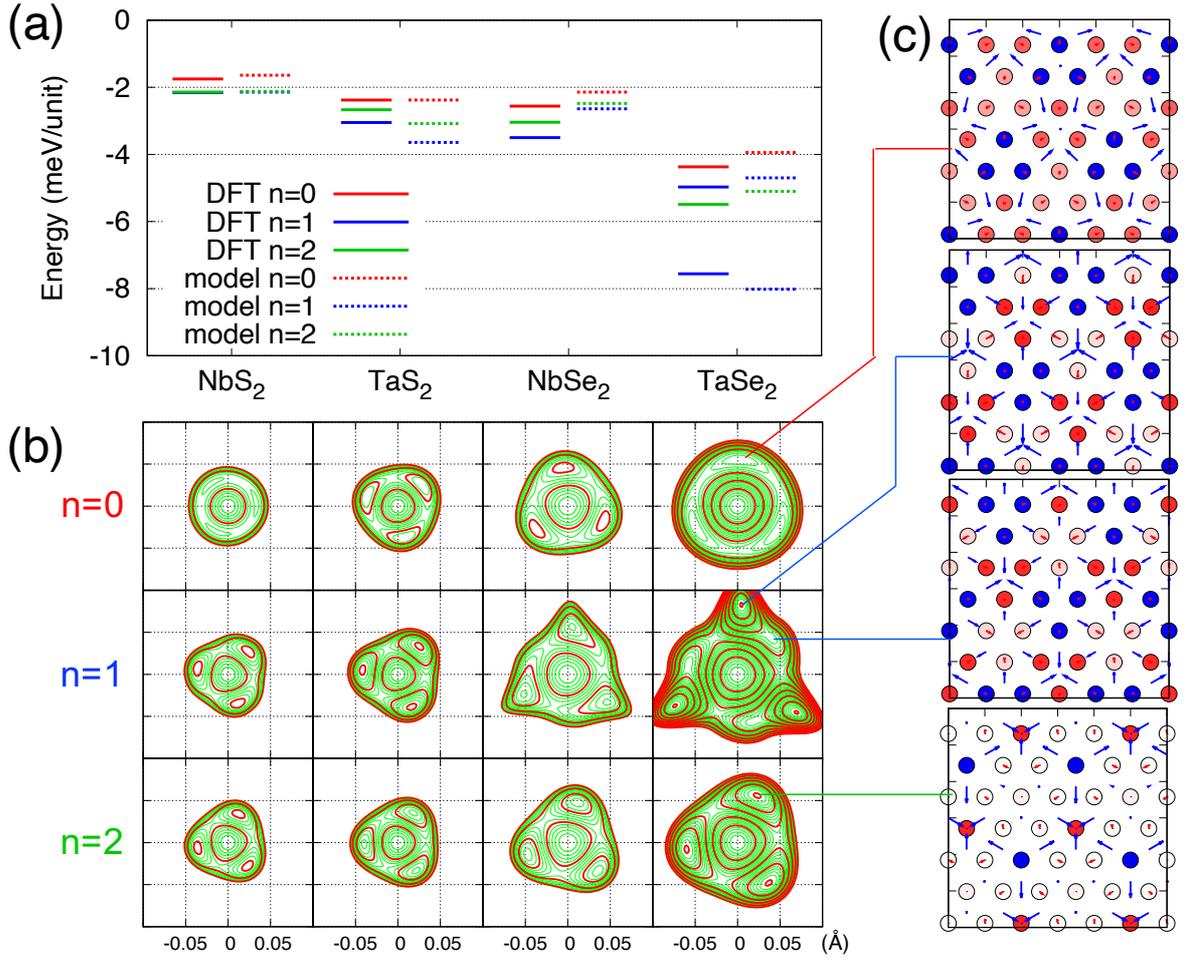

FIG. 2. (a) Formation energies of commensurate CDWs for monolayer H-NbS$_2$, H-TaS$_2$, H-NbSe$_2$, and H-TaSe$_2$. Solid lines correspond to density functional theory results and dashed lines are from Eq. (1). Red, blue and green colors denote three distinctive CCDWs $n = 0$, 1 and 2, respectively. (b) Contour plots of potential landscape $E_n(A, \theta)$ in polar coordinate. The energy spacing of red (green) contour is 1.0 (0.2) meV per unitcell and $E_n(0,0) = 0$ meV. (c) The displacement vectors of four local minimum CCDW structures of monolayer H-TaSe$_2$. Blue arrows indicate the displacements of Ta atoms and circles are Se atoms of upper plane. Out-of-plane displacements of Se atoms are color-coded with reddish (bluish) color for upward (downward) direction.

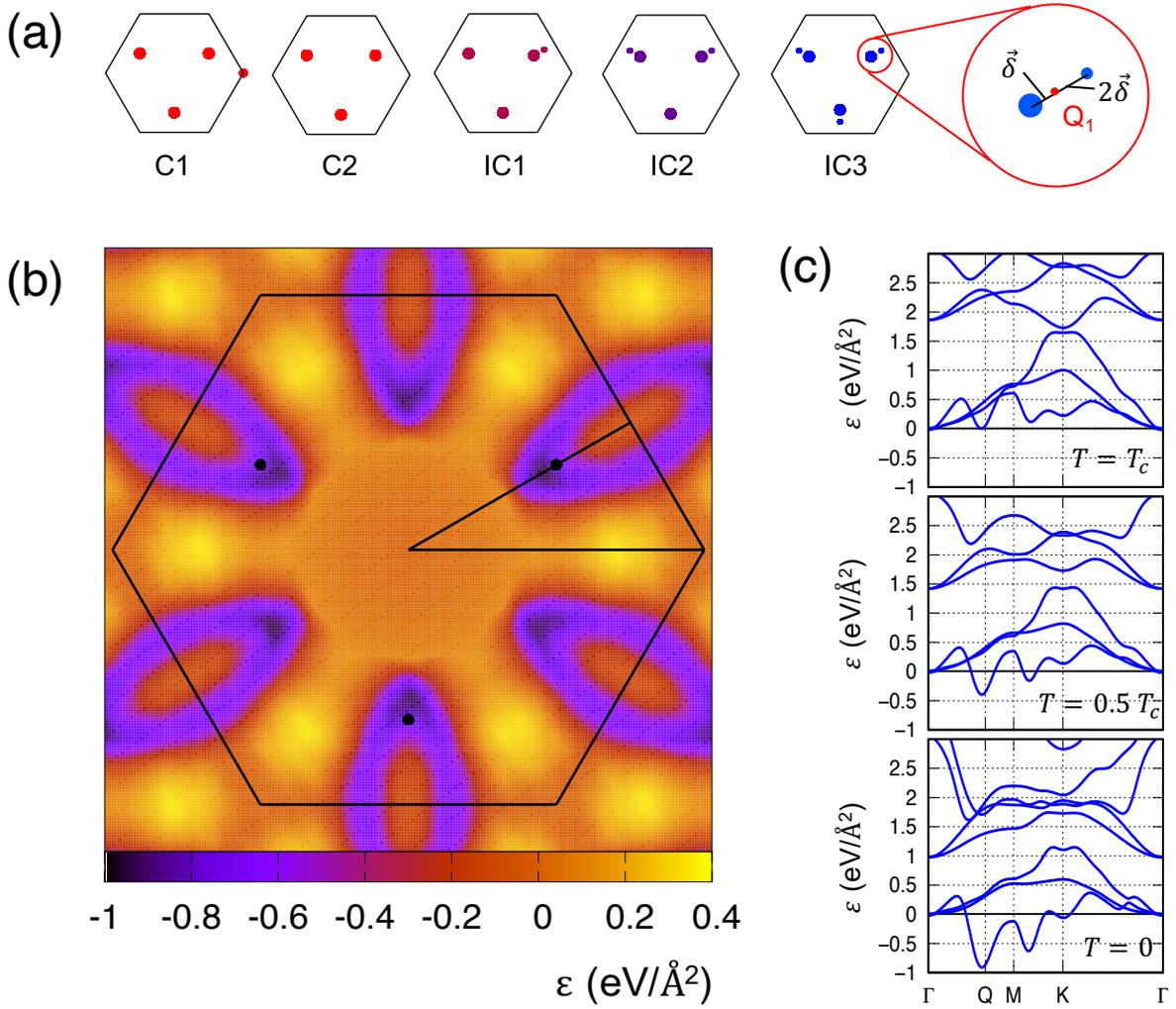

FIG. 3. (a) Five variational CDWs for commensurate (C1 and C2) and incommensurate (IC1, IC2 and IC3) phases. The wavevectors of main eigenmodes (large dots) reside around three $Q$'s and $K$. Whenever they slightly deviate from $Q$ by $\vec{\delta}$, eigenmodes with wavevector (small dots) $\vec{Q} - 2\vec{\delta}$, accompany them. Optimizations are performed for amplitudes and phases of commensurate phase, while for incommensurate phases, $\vec{\delta}$ is also optimized. Color map in (b) denotes $\varepsilon(\vec{k})$ of lowest-branch eigenmodes of monolayer H-TaSe$_2$ at $T = 0$. $\varepsilon(\vec{k})$ along high-symmetric lines are shown in (c) for $T = 0$, 0.5, and 1.0 $T_c$, respectively.

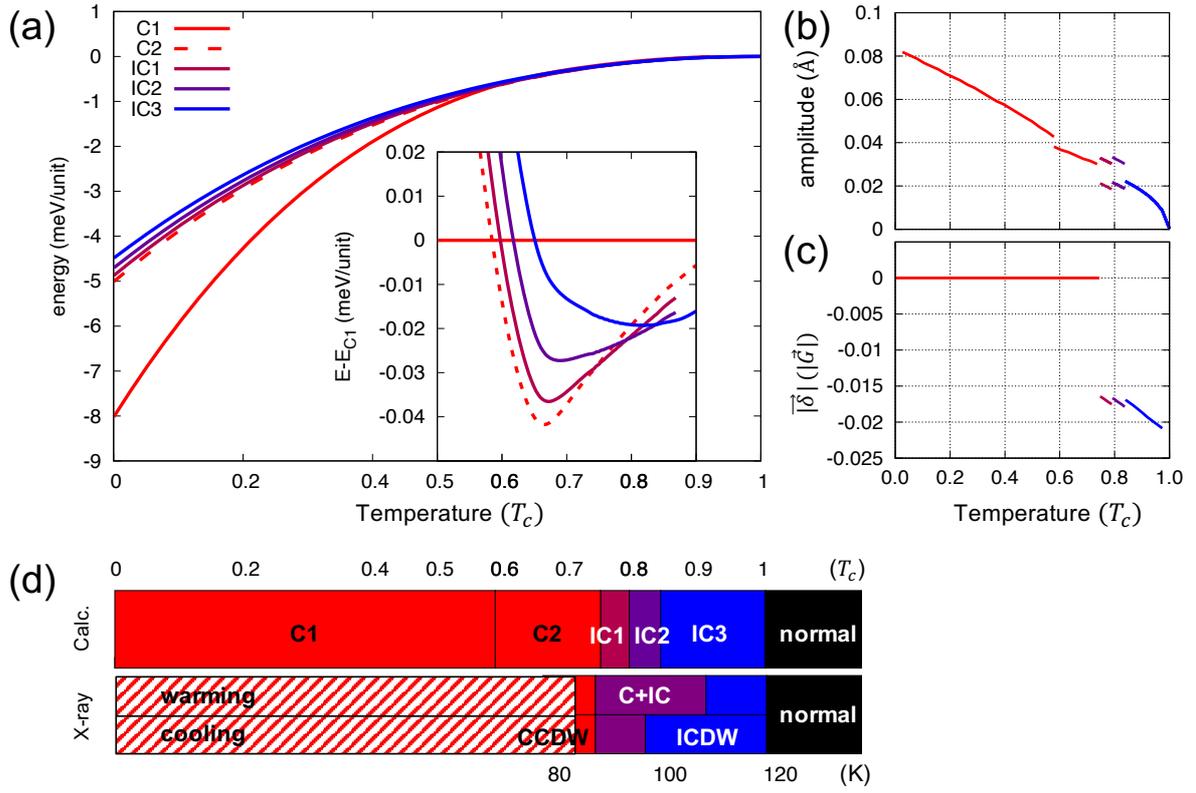

FIG. 4. (a) The energies, (b) amplitudes of eigenmodes, and (c) deviations $|\vec{\delta}|$ of five CDW in monolayer H-TaSe$_2$ as a function of temperature. Inset in (a) is the magnified view of the energy difference between phases. (d) Calculated phase diagram scaled by $T_c$ (upper) and the experimental phase diagram (lower) of bulk 2H-TaSe$_2$ with X-ray diffraction measurements in Ref. 31. The hatched region ($T <$ 85 K) is not measured.

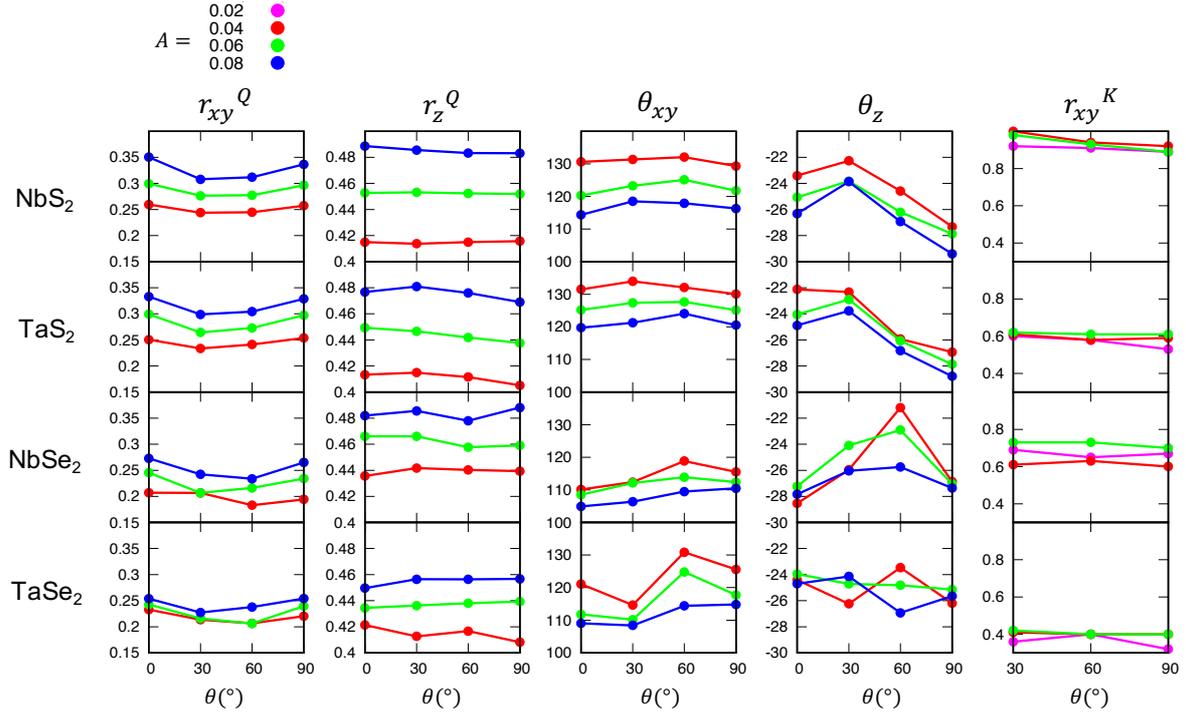

FIG. 5. For monolayer H-TMDC's, parameters of polarization vectors are calculated for $Q$ (left four columns) and $K$ eigenmodes (right column) with given amplitudes (colors) and phases ($x$-axes).

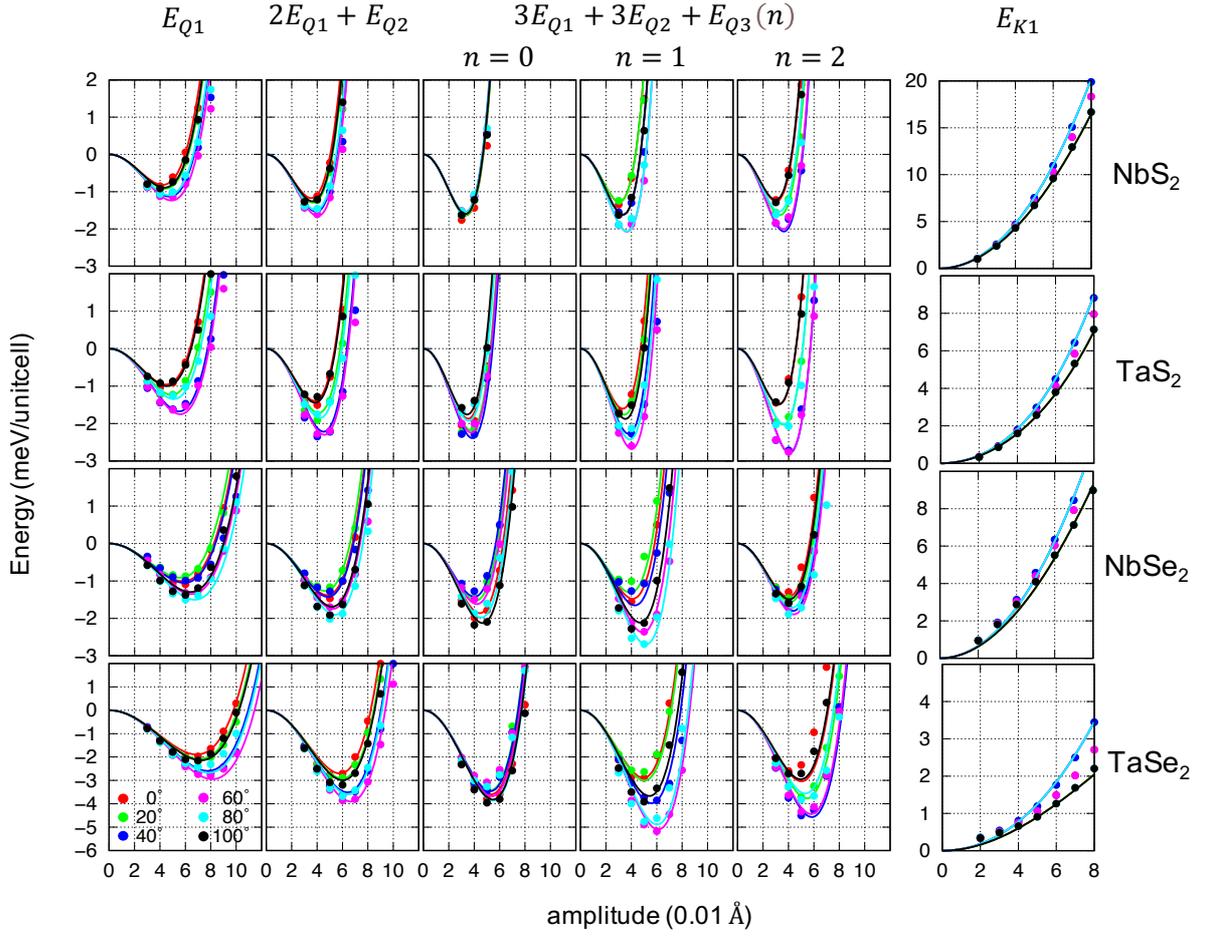

FIG. 6. Least square fittings of configurations only with $Q$ eigenmodes (the first five columns) and $K$ eigenmodes (right column). The form of fitting function (line) is on the top of each column for the corresponding configurations (points). The amplitudes and phases are represented with $x$-axes and colors.

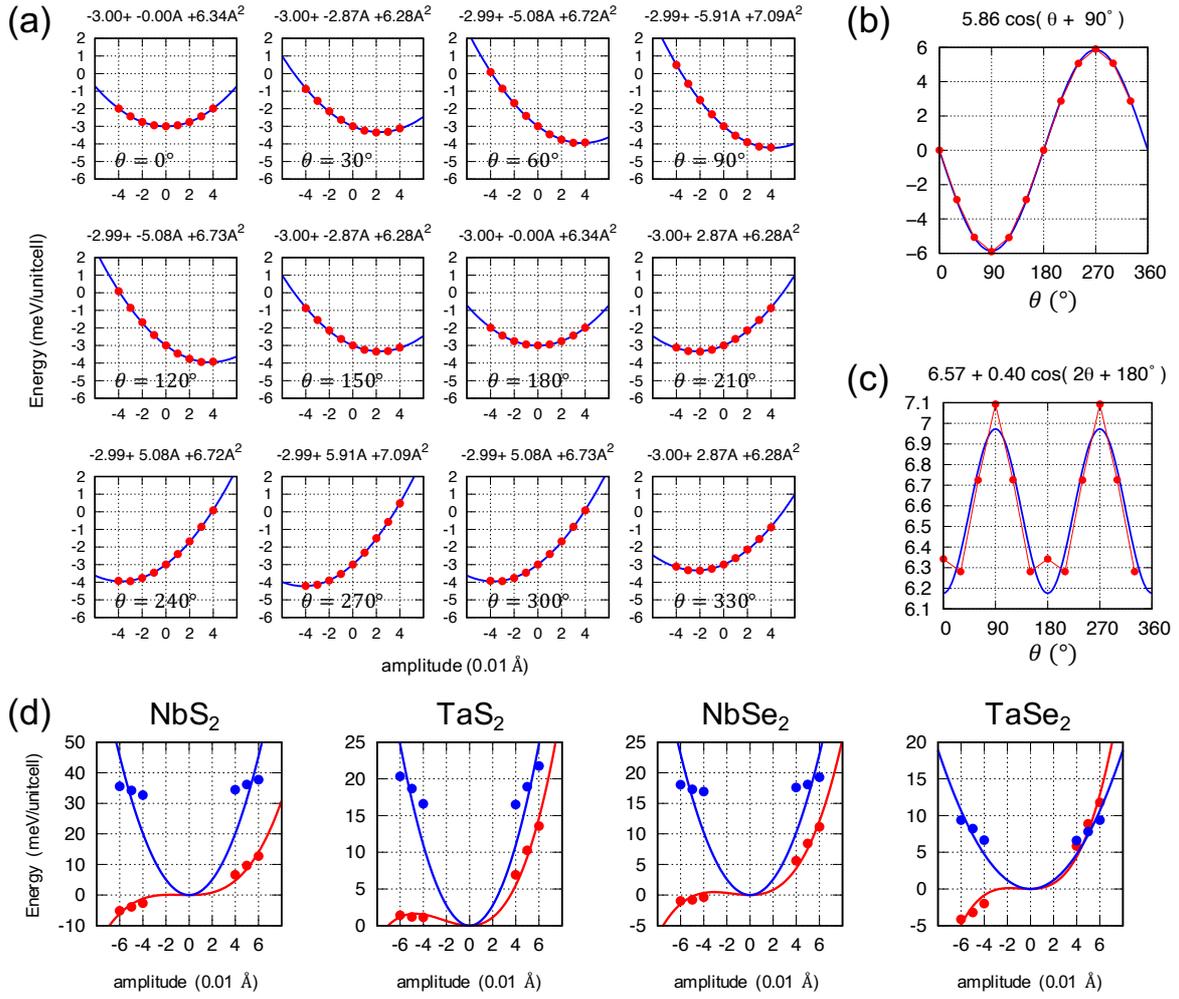

FIG. 7. (a) Energies (red points) of a specific $n = 1$ configuration (see text) superposed with a $K$ eigenmode. Each panel corresponds to the phase of $K$ eigenmode for various amplitudes ($x$-axes). The title of each panels is the equation of second order polynomial fittings function (blue line). The unit of $A$ is 0.1 Å. First (b) and second (c) order coefficients of (a) are plotted for the phase of $K$ eigenmode. Titles are fitting functions for them. Amplitude of cosine functions in (b) is $3c_{QK}A_Q^2 + 3d_{QK1}A_Q^3$ and the constant in (c) is $3D_{QK}A_Q^2$. For various $n = 1$ configurations with amplitudes $A_Q$ ($x$-axes), the formers (latters) are plotted in red (blue) points in (d). Lines are least-square fitting results with coefficients $c_{QK}, d_{QK1}$ and $D_{QK}$.

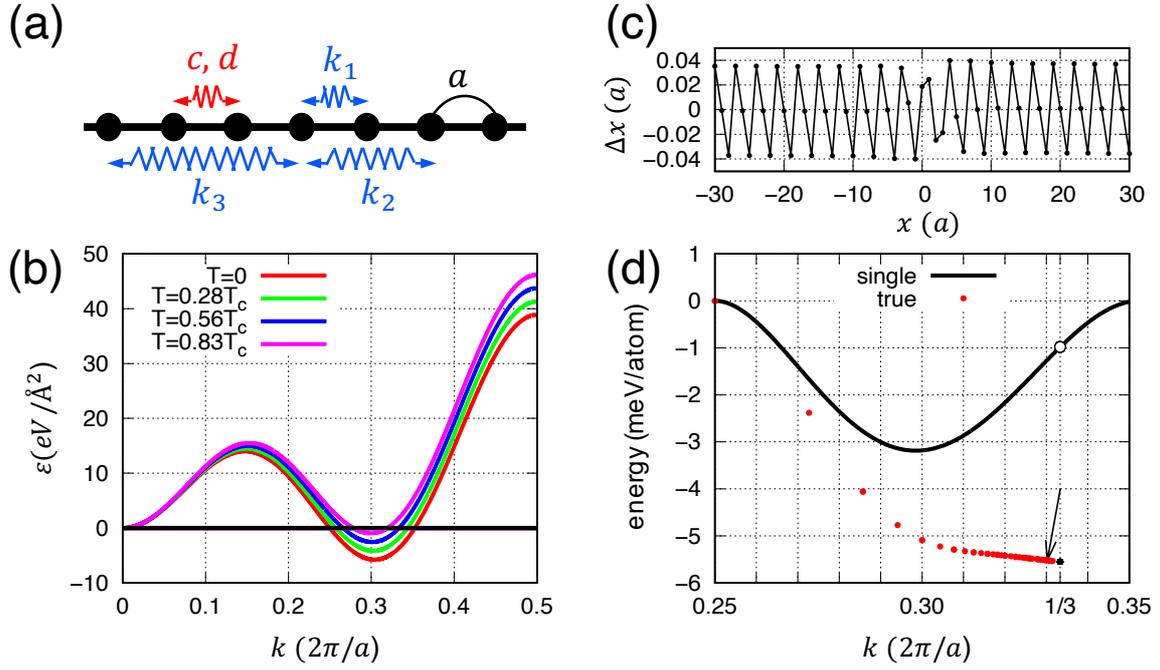

FIG. 8. (a) Atomic chain with nearest third- (fourth-) order force constant $c$ ($d$) and $i$-th neighbor second-order force constants $k_i$'s. (b) The second-order potential energy $\varepsilon$ of an eigenmode with wavevector $k$ and unit amplitude 1 Å (red) for $k_1 = 0.3, k_2 = -0.4$ and $k_3 = 0.5$ Ry/bohr². At $T = T_c$, all eigenmodes have positive potential energies. c, A real space atomic displacements of incommensurate structure indicated by an arrow in (d). (d) Minimum potential energy of a single eigenmode with $k$ (black) and true ground states (red) under the constraint of a periodic boundary condition allowing $k$.

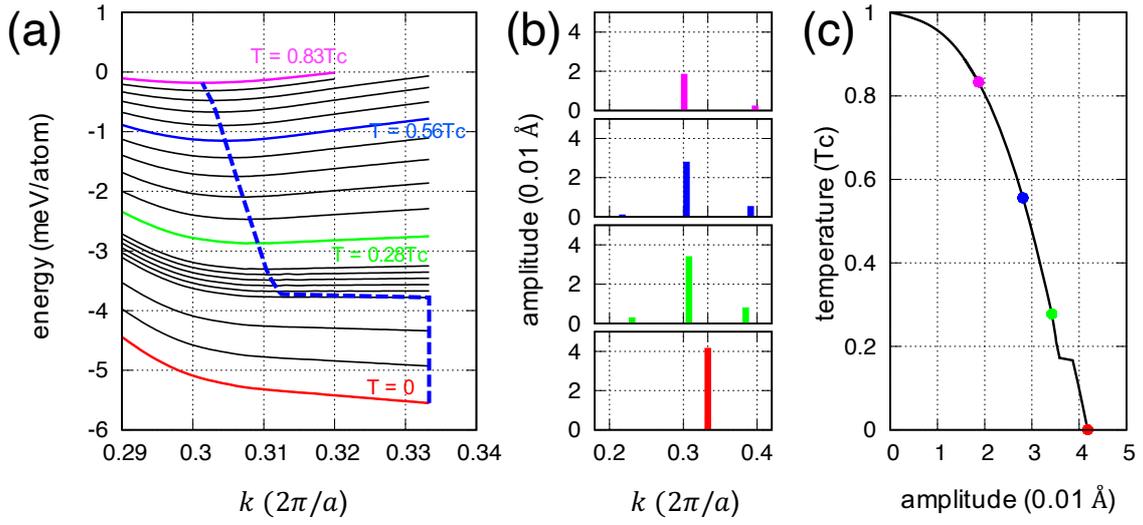

FIG. 9. (a) Ground states energies under the constraint of a periodic boundary condition allowing $k$ for finite temperatures. Temperatures are chosen in an arbitrary spacing, increasing from the bottom to the top. Dashed blue line is an eye guide for the global minima depicting period change in the modulation on temperature. (b) Fourier components of global minimum structures at selected temperatures as color-indicated in (a). (c) Temperature-dependence of the largest Fourier components of global minimum structures (solid line). Four selected temperature in (a) are indicated by points with corresponding colors.